\documentclass{article}
\usepackage{spconf,amsmath,epsfig}
\usepackage{url}

\usepackage{mathtools}
\usepackage{amsfonts}
\usepackage{amssymb}
\usepackage{multirow}
\usepackage{multicol}
\usepackage{booktabs}
\usepackage{subfigure}

\title{
FEEDBACK Neural Network based Super-resolution of  DEM for generating high fidelity features}
\name{Ashish A Kubade, Avinash Sharma, K S Rajan}
\address{International Institute of Information Technology, Hyderabad}
\begin{document}
%
\maketitle

\begin{abstract}
High resolution Digital Elevation Models(DEMs) are an important requirement for many applications like modelling water flow, landslides, avalanches etc. Yet publicly available DEMs have low resolution for most parts of the world. Despite tremendous success in image super resolution task using deep learning solutions, there are very few works that have used these powerful systems on DEMs to generate HRDEMs. Motivated from feedback neural networks, we propose a novel neural network architecture that learns to add high frequency details iteratively to low resolution DEM, turning it into a high resolution DEM without compromising its fidelity. Our experiments confirm that without any additional modality such as aerial images(RGB), our network DSRFB achieves RMSEs of 0.59 to 1.27 across 4 different datasets.
\end{abstract}
\begin{keywords}
Digital Elevation Models, Terrains, Super-resolution, Feedback Neural Networks
\end{keywords}

\footnote{Copyright 2020 IEEE. Published in the IEEE 2020 International Geoscience \& Remote Sensing Symposium (IGARSS 2020), scheduled for July 19 – 24, 2020 in Waikoloa, Hawaii, USA. Personal use of this material is permitted. However, permission to reprint/republish this material for advertising or promotional purposes or for creating new collective works for resale or redistribution to servers or lists, or to reuse any copyrighted component of this work in other works, must be obtained from the IEEE. Contact: Manager, Copyrights and Permissions / IEEE Service Center / 445 Hoes Lane / P.O. Box 1331 / Piscataway, NJ 08855-1331, USA. Telephone: + Intl. 908-562-3966.}
\section{Introduction}
\label{sec:intro}
Digital Elevation Models (DEMs) are of paramount importance in Geographical Information systems(GIS) as they enable accurate modeling of terrain features and provide the basis for studying dynamic phenomena like erosion, land slides, flooding, etc. In addition, many engineering applications like visibility analysis, cut-and-fill problem, landscape designing also require the knowledge of the topographic features and surface variations. Hence, there is a need for a high resolution DEM with a near-true representation of the terrain features. 
The majority of publicly available DEM datasets like SRTM, ASTER, CartoDEM are of relatively lower spatial resolution (~30 meters), except for selected geographical locations where more finer spatial resolution data is available ($<2$ meters, e.g., OpenDEM~\cite{opendem}). Recently, airborne LiDAR technology has enabled capturing high resolution DEMs, yet getting LiDAR scans and processing the data is quite time consuming and expensive. Thus, it is important to explore scalable solutions for generating High Resolution DEMs (HRDEMs) using their low resolution counterparts i.e., LRDEMs. 
%
\begin{figure}[t]
\includegraphics[width=0.48\textwidth]{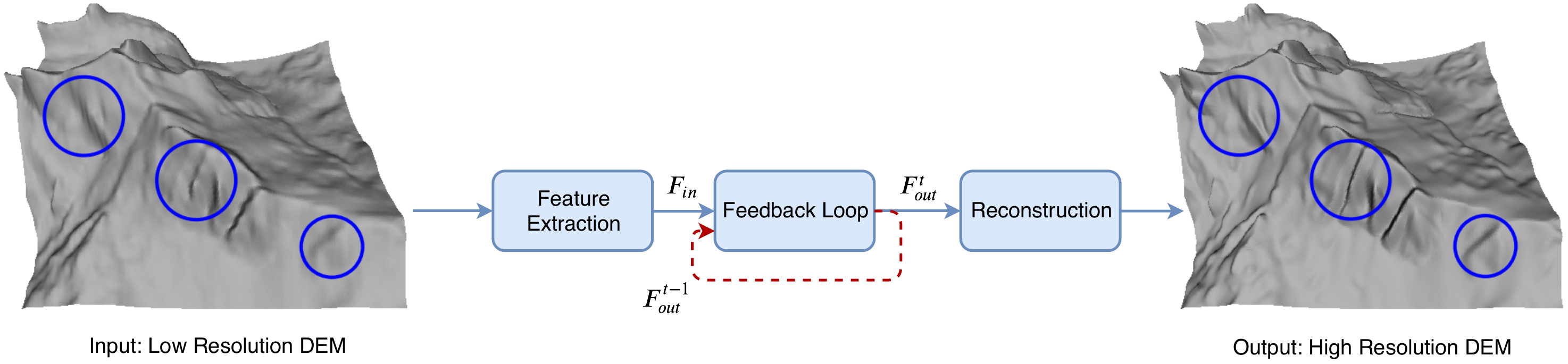}
\centering
\caption{Outline of the proposed DSRFB network that transform input low resolution DEM to high resolution DEM. Blue circles depict the recovered high fidelity features.}
\label{fig:intro}
\vspace{-0.5cm}
\end{figure}

Existing methods for generating HRDEMs, also known as terrain amplification, have primarily followed two paradigms: one aimed at reconstructing DEMs with higher fidelity with respect to the actual terrain (e.g., dictionary based reconstruction~\cite{argudo2017coherent}), and other focused on adding plausible details and enhancing visual appearance, apathetic to deviation from real terrains, e.g., ~\cite{guerin2017interactive}. 
Recently,~\cite{argudo2018terrain} proposed a fully convolutional network (FCN) architecture that uses LRDEM along with registered ortho-photo (RGB) to generate an HRDEM. However, generating registered ortho-photo and LRDEM pairs is an arduous task. Additionally, such image modality could be misleading due to dense vegetation, heavy snow and timely changes in the landscape appearance. 

In this work, we propose a novel supervised approach to super-resolve the DEMs from LRDEMs. In DEMs majority of the higher frequency details are available as small discontinuous edges(fractal structures) unlike natural images which often contain larger structures. This motivates us to focus on learning features at smaller scales. Interestingly, the initial layers in CNN capture these smaller scale (high frequency) details, which need to be enhanced or strengthened for DEM super-resolution. Recently, \cite{li2019feedback} proposed a feedback module in their super-resolution architecture (called SRFBN) for RGB images to refine the features learned by initial layers.
%
Hence, to specifically enhance the  lower level (initial layer) DEM features, we propose to use the similar feedback module in our architecture. As shown in Fig.\ref{sec:intro}, the features extracted from LRDEM i.e. $F_{in}$ are processed along with successive outputs of Feedback block i.e. $F^{t-1}_{out}$ to generate finer features as $F^t_{out}$. Hereinafter, we call the proposed network as DSRFB (DEM Super-Resolution based on Feedback Network) 
The results from DSRFB using only LRDEM as input shows improved results in comparison to the state-of-the-art approach using LRDEMs combined with orthoimagery. 
In short, our contributions are as follows:
\begin{itemize}
\item 
A new Super-resolution network for DEMs that iteratively refines lower level features in DEM for better fidelity 
\item
The proposed method overcomes the feature split across tile boundaries by incorporating multiple estimates in HRDEM reconstruction
 \end{itemize}
\section{Related Work}
We briefly introduce the most relevant work in super-resolution  using  Convolutional Neural Networks(CNNs).
Increasing the resolution of a DEM can be seen as hallucinating the higher frequency details which are not present in LRDEM at all. Bicubic and bilinear interpolation are the simplest methods for finding such details. Although these methods often end up smoothing the texture, these can be used as an initial step towards super-resolution.
In images, Dong Et al\cite{dong2014learning}, used bicubic interpolation first to upsample LR and then used 3 layered CNN, with mean squared error over pixels as the loss function to generate better HR images.
Further \cite{dong2016accelerating} allowed network itself to learn upsampling kernels so as to bypass the bicubic step for real time applications.
Availing the power of more advanced Generative Adversarial Networks(GANs), where two neural networks, Generator and Discriminator, adversarily try to beat each other's performance, Johnson et al\cite{johnson2016perceptual} achieved visually appealing results in images.
For terrain visualization, \cite{guerin2017interactive} adopted GAN based style transfer using sketch cues over DEMs. 
Using Recurrent Neural Networks(RNNs), authors of \cite{li2019feedback} proposed to use a feedback module to iteratively refine lower level features in images. We find use of such a feedback module can be helpful to correct the lower features in terrains as well.
Oscar et al\cite{argudo2018terrain} used Fully Convolutional Network(FCN) architecture for super-resolution of DEMs. Their method uses a georegisted RGB image along with LRDEM to predict the HRDEM. While their two branch FCN architecture gives state-of-the art performance, usage of their method is limited by availability of pair of georegistered DEM and aerial image.
\section{Method}
We treat super-resolution as a Image translation problem. As in \cite{dong2014learning}, we upsample LRDEM with bicubic interpolation as a preprocessing step and denote it as ILR and then add higher details by using the proposed network. Our network was inspired from SRFBN\cite{li2019feedback} where we wish to learn more effective low level features by receiving feedback from  higher level features.
\subsection{Network Architecture}
\begin{figure}[t]
\includegraphics[width=0.5\textwidth]{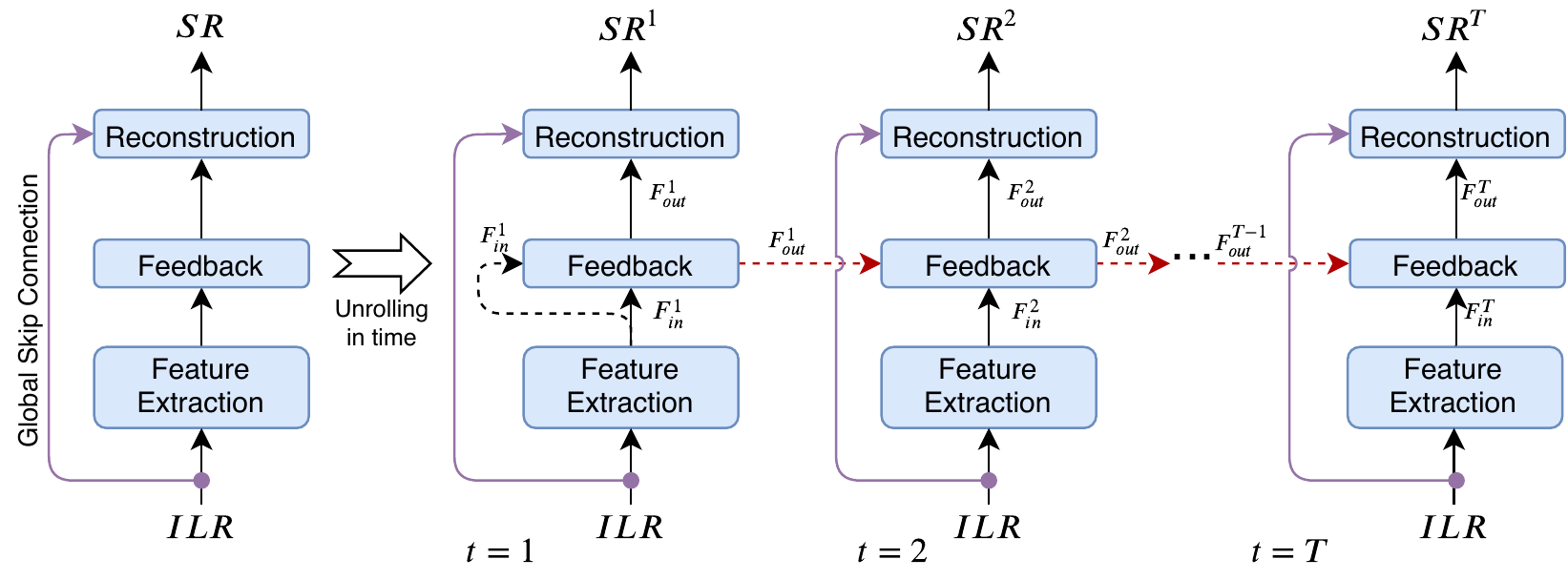}    
\centering
\caption{Network unrolled in time steps}
\label{unrolled}
\vspace{-0.5cm}
\end{figure}

Our main task is to propagate higher level feature information to lower levels. Our proposed network achieves this using a recurrent neural network(RNN). An RNN is a neural network with cyclic connections that make them capable of handling sequential data. An RNN with $T$ hidden states can store network activation up to T states. We model our feedback network such that with each time step, the unrolled network reconstructs a new SR image and also store the layer activations in hidden state which are used for the next time step.  Fig. \ref{unrolled} shows, the network unfolded in time steps. Each sub-network consists of three parts: Feature Extraction(FE) block to capture low level details from ILR image, followed by a Feedback module(FB) where we refine the low level features using high level features and lastly a Reconstruction block(RB).
The Feature Extraction(FE) module consists of two convolution layers Conv(\textit{m},3) and Conv(\textit{$4 * m$},1), where \textit{m} is the base number of filters.
DEMs up-sampled with bicubic interpolation(ILR) are used as input for feature extraction. FE module encodes ILR, denoted as \textbf{${F^t_{in}}$} which is then forwarded to Feedback(FB) module once for each time step $t \in \{1,T\}$. 

\begin{figure}[t]
\includegraphics[width=0.5\textwidth]{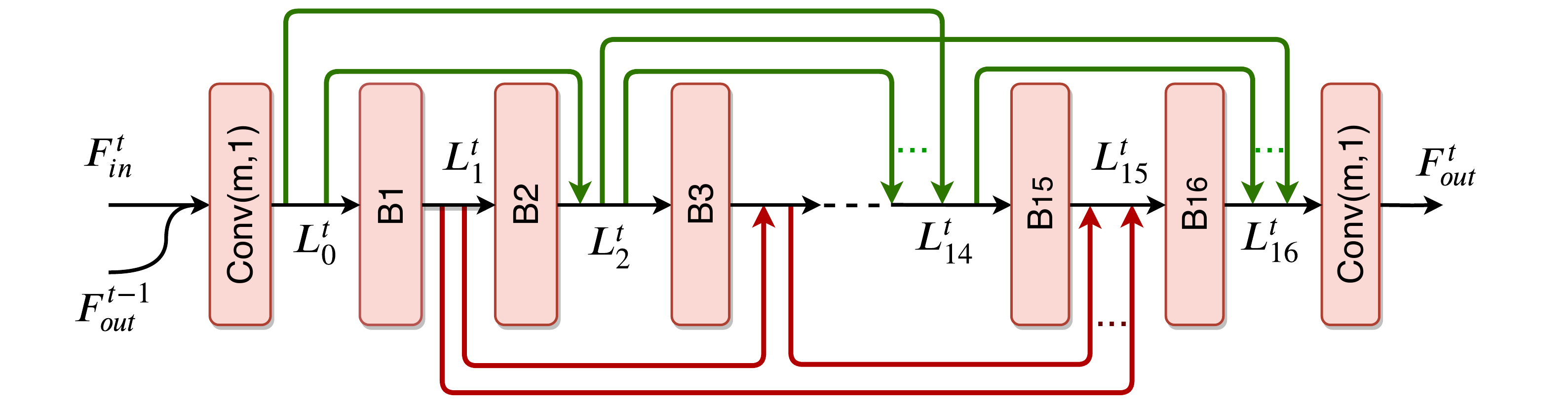}    
\centering
\caption{Layers and skip connections in Feedback Module.}
\label{FB_module}
\vspace{-0.5cm}
\end{figure}

Our Feedback(FB) module consists of stack of residual units.
Each residual unit consists of a Conv(m,1) followed by a Conv(m,3) layer. The purpose of Conv(m,1) layers is to adaptively fuse the number of input channels to base number of filters. We denote the residual units as \textbf{$B_{i}$} where $i$ $\in$ \{1,n\}.  We use n=16 in our case. At a particular time step `t', the output of residual unit \textbf{$B_{i}$} is denoted as \textbf\textit{$L_{i}^{t}$}.  At time step `t', the state of our feedback module is shown in Fig. \ref{FB_module}. Inspired from DRRN\cite{tai2017image}, we use multi-path skip connections with a few changes. Instead of using the residual connections from a current residual unit to all its subsequent units, we use them in alternate subsequent units. 
Thus the information $L^t_{1}$ would be passed to $\{B2, B4,..., B16\}$, $L_2^{t}$ to $\{B3,B5,...,B15\}$, $L_3^{t}$ to $\{B4,B6,...,B16\}$ and so on. Eventually, we get two groups of residual paths as shown by red and green arrows in Fig. \ref{FB_module}. The feedback module receives a concatenation of $F^{t-1}_{out}$ and $F^{t}_{in}$ as input, denoted as [$F^{t-1}_{out}, F^{t}_{in}$], where $F^{t-1}_{out}$ is the hidden state output features from previous time step `t-1' and $F_{in}^{t}$ is the input feature vector at time step `t'. We use Conv(m,1) to compress [$F^{t-1}_{out}, F^{t}_{in}$] along the channels and denote the compression output as $L^{t}_{0}$. We also use residual connections for $L^{t}_{0}$ to all the alternate layers except the last layer.


Lastly, the outputs of $\{B2,B4,...,B16\}$ are compressed by a Conv(m,1) into $F^t_{out}$, which gets forwarded to reconstruction block for SRDEM generation and also used as feedback information for the next iteration which we concatenate with $F^{t+1}_{in}$. As there will be no feedback available at iteration t=1, we use $F^1_{in}$ as $F^0_{out}$.

So, at the start of each iteration, generalized input will be [$F^{t}_{in}, F^{t-1}_{out}$] which will be compressed by the Conv(m,1).
Inside Reconstruction block(RB) the output of feedback module $F^{t}_{out}$ at iteration t is operated with Conv(1,3) layer to generate a residual DEM image $I_{Res}^{t}$. Using the global residual learning, we pass the interpolated LRDEM denoted as ILR via a global skip connection. The SR image at iteration $t$ is then reconstructed as,
 \begin{equation}
     I^{t}_{SR} = I_{Res}^{t} + \text{ILR} 
 \end{equation}
Finally, for $T$ number of hidden states, the model will generate a collection of \textit{T} SR images ($I^{1}_{SR}, I^{2}_{SR},..., I^{T}_{SR}$). For each of the time step, we evaluate \textbf{\textit{L1 loss}} (mean absolute error) between generated SR DEM and ground truth HRDEM.
\begin{equation}
\label{l1_loss}
    loss = \sum_{i=1}^{T} | I^t_{SR} - \text{HRDEM} |
\end{equation}
As shown in equ. \eqref{l1_loss}, we accumulate losses over all T steps and at the end of T steps, we back-propagate the loss to optimize the parameters of the network.
\section{Experiments and Results}
\subsection{Datasets}
We use the same dataset used by \cite{argudo2018terrain} which is part of publicly available high resolution DEMs datasets named Pyrenees~\cite{icc} and Tyrol~\cite{sbg}, respectively. The data is in the form of tiles with a spatial resolution of 2m. 
The data has been organized in create LR-HR pairs by down sampling HR tiles to a spatial resolution of 15m. We explicitly mention that, with a LR-HR pair of resolutions 15m-2m, the effective scale of super-resolution in our experiments will be 7.5x rather than 8x. 
For fair comparison, we use same distribution as used by \cite{argudo2018terrain} as combining tiles from both Pyrenees and Tyrol, with 22000 training and 11000 validation tiles, with each tile of 200x200 pixels cropped from the original larger tiles. Four regions named Bassiero, Forcanada, Dürrenstein and Monte Magro have been set for testing.
\subsection{Implementation Details}

The kernel size for all convolution layers is 3x3 except for the compression layers where we use 1x1 kernels. PReLU follows all the Conv layers, except in the reconstruction block. We choose SR factor as 8X, m as 64, T as 4 and n as 16. We use SR image at t=4 as the final SR image. We also consider ensemble of SR images for all iteration from t=1 to t=4, to get a marginal improvement over the reconstruction. We explore two ways to reconstruct the HRDEM patches from the network. With simplest setting in DSRFB, we reconstruct each patch independently and place them in the larger tile. However, to effectively recover the features split across the boundary regions we propose DSRFO to process patches with an overlap. We use an overlap of 25\% on each side and use aggregated response for the pixels in the overlapped region.
We use the entire $200$x$200$ tile as a patch for training. We experimented with smaller patch sizes as well and found that because of boundary regions, the larger the patch size the better the outcome. We use batch size of 4, \textit{adam} optimizer with learning rate of $0.0001$, and weights initialized with \textit{kaiming} initialization. We use multi-step learning degradation with gamma set to $0.5$. Pytorch was used for implementation of the network and it was trained for 100 epochs on NVIDIA 1080Ti GPUs.

\begin{table*}[]
\centering
\begin{tabular}{|c@{\hskip2pt}|c|c|c|c|c@{\hskip2pt}|c|c|c|c|c@{\hskip2pt}|}
\hline
\multirow{4}{*}{Region} & \multicolumn{5}{c|}{PSNR (in dB, the higher the better)} & \multicolumn{5}{c|}{RMSE (in meters, the lower the better)} \\ \cline{2-11} 
 & \multirow{2}{*}{Bicubic} & \multirow{2}{*}{FCND} & \multicolumn{2}{c|}{Ours} & FCN & \multirow{2}{*}{Bicubic} & \multirow{2}{*}{FCND} & \multicolumn{2}{c|}{Ours} & FCN \\ \cline{4-5} \cline{9-10}
 &  &  & DSRFB & DSRFO & (+ RGB) &  &  & DSRFB & DSRFO & (+ RGB) \\ \hline
Bassiero & 60.5 & 62.261 & 62.687 & \textbf{62.752} & 63.4 & 1.406 & 1.146 & 1.091 & \textbf{1.083} & 1.005 \\ \hline
Forcanada & 58.6 & 60.383 & 60.761 & \textbf{60.837} & 62.0 & 1.632 & 1.326 & 1.2702 & \textbf{1.259} & 1.097 \\ \hline
Dürrenstein & 59.5 & 63.076 & 63.766 & \textbf{63.924} & 63.6 & 1.445 & 0.957 & 0.884 & \textbf{0.868} & 0.901 \\ \hline
Monte Magro & 67.2 & 70.461 & 71.081 & \textbf{71.196} & 71.1 & 0.917 & 0.632 & 0.589 & \textbf{0.581} & 0.587 \\ \hline
\end{tabular}
\caption{PSNR and RMSE for the test regions. Note that our networks (DSRFB and DSRFO) do not use additional RGB image and still outperform other state-of-the-art methods (including FCN(+ RGB) for some of the regions).}
\label{rmse_res}
\end{table*}

\begin{figure*}[htb]
\begin{minipage}[b]{0.16\linewidth}
  \centering
\centerline{\epsfig{figure=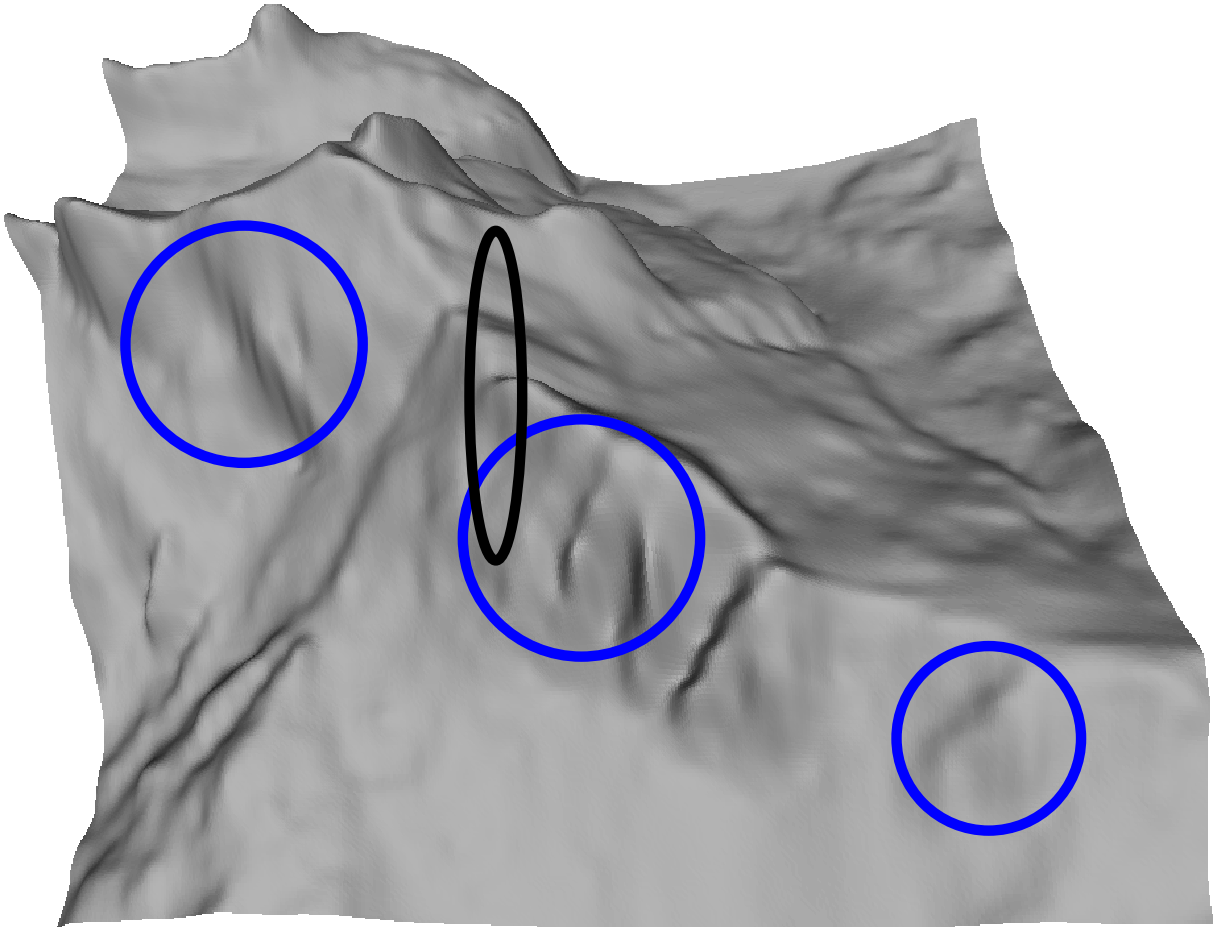,width=2.8cm}}
  \centerline{Bicubic}\medskip
\end{minipage}
\begin{minipage}[b]{0.16\linewidth}
  \centering
\centerline{\epsfig{figure=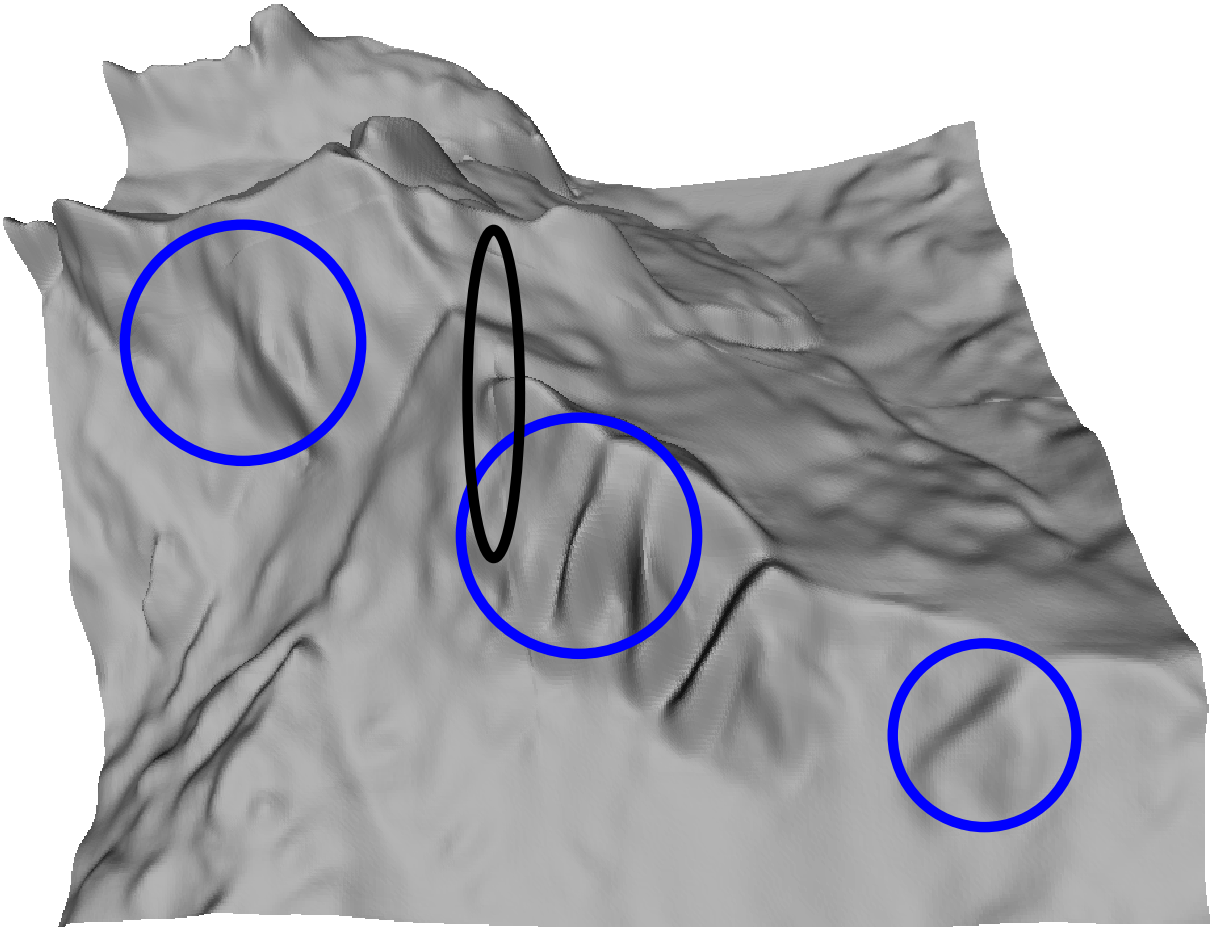,width=2.8cm}}
  \centerline{FCND}\medskip
\end{minipage}
\begin{minipage}[b]{.16\linewidth}
  \centering
\centerline{\epsfig{figure=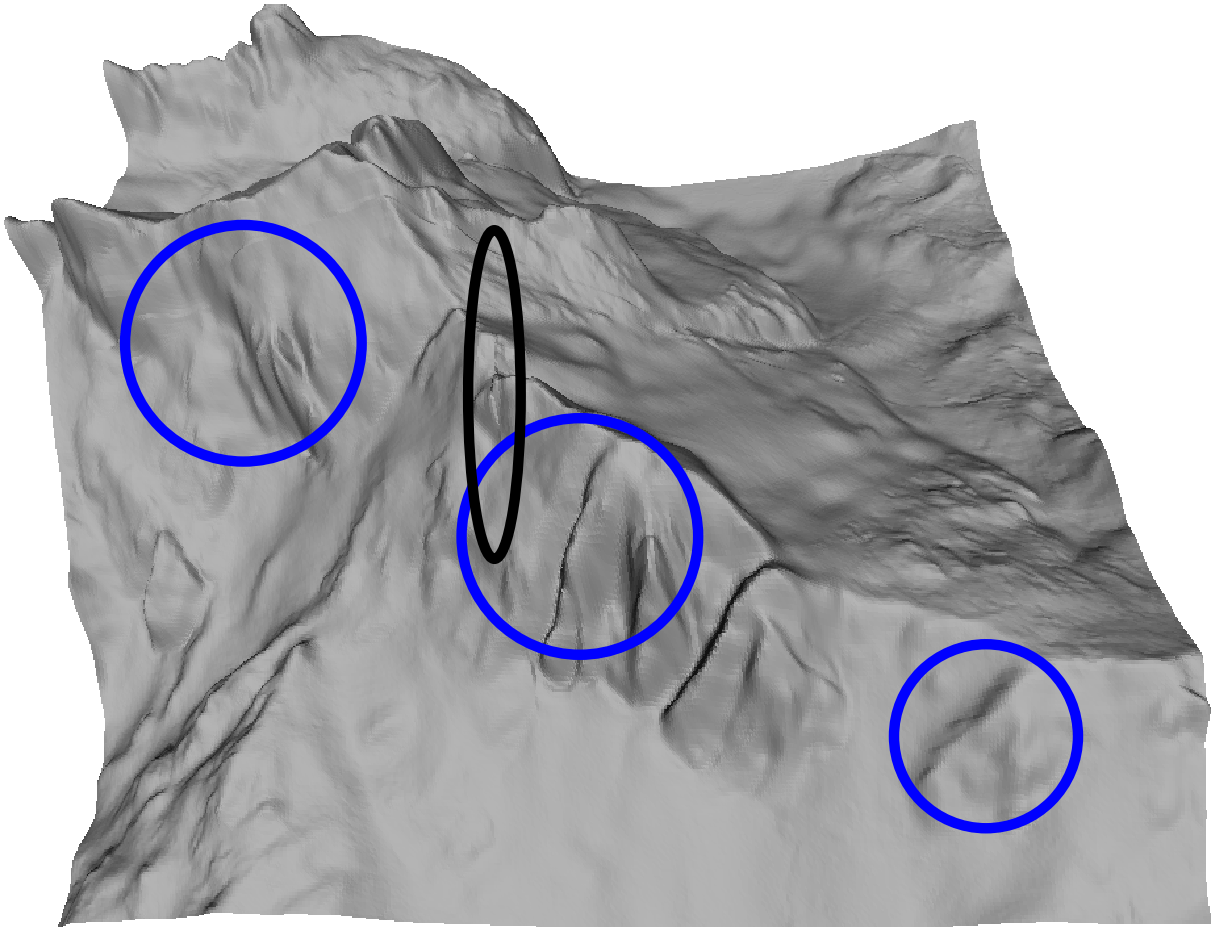,width=2.8cm}}
  \centerline{FCN}\medskip
\end{minipage}
\begin{minipage}[b]{.16\linewidth}
  \centering
\centerline{\epsfig{figure=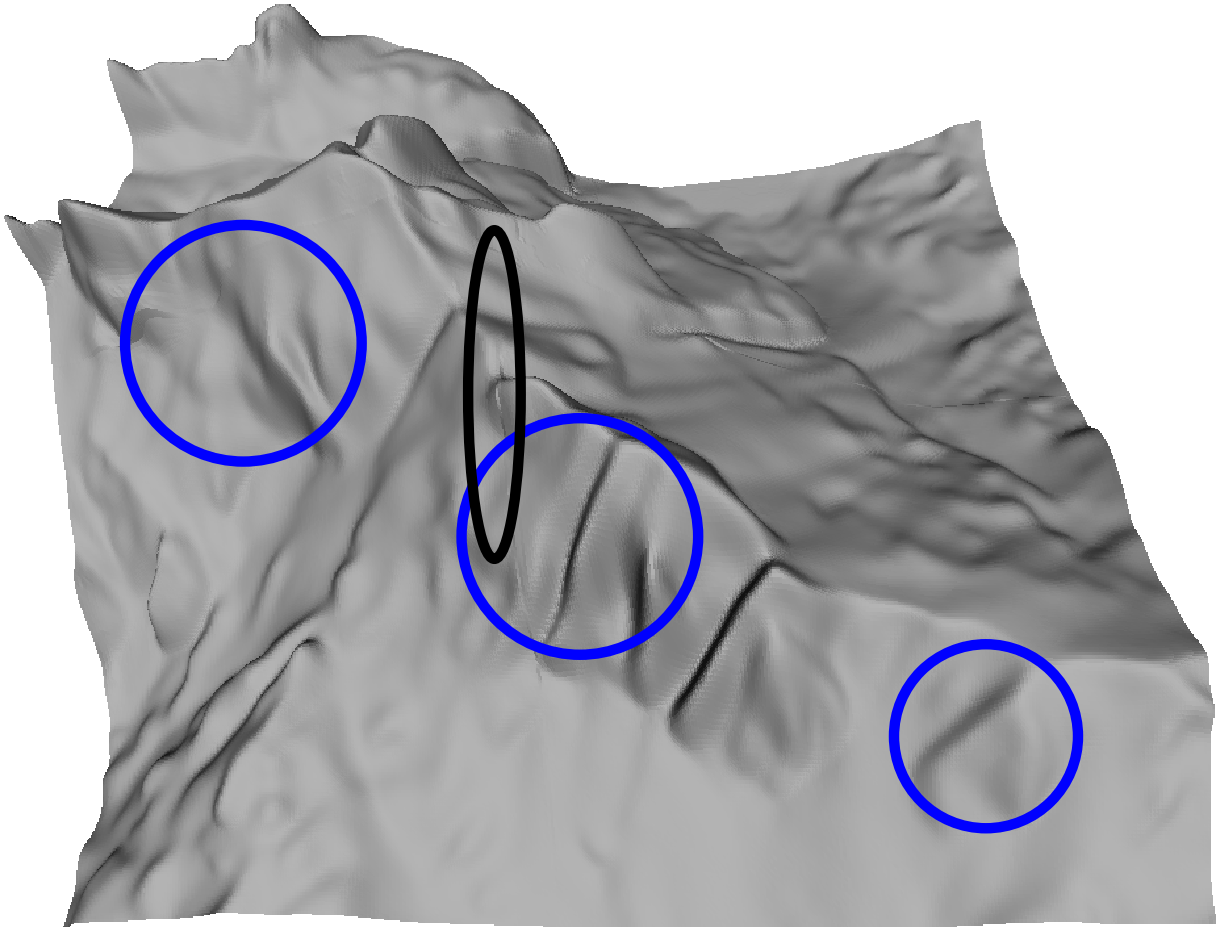,width=2.8cm}}
  \centerline{DSRFB (Ours)}\medskip
\end{minipage}
\begin{minipage}[b]{0.16\linewidth}
  \centering
\centerline{\epsfig{figure=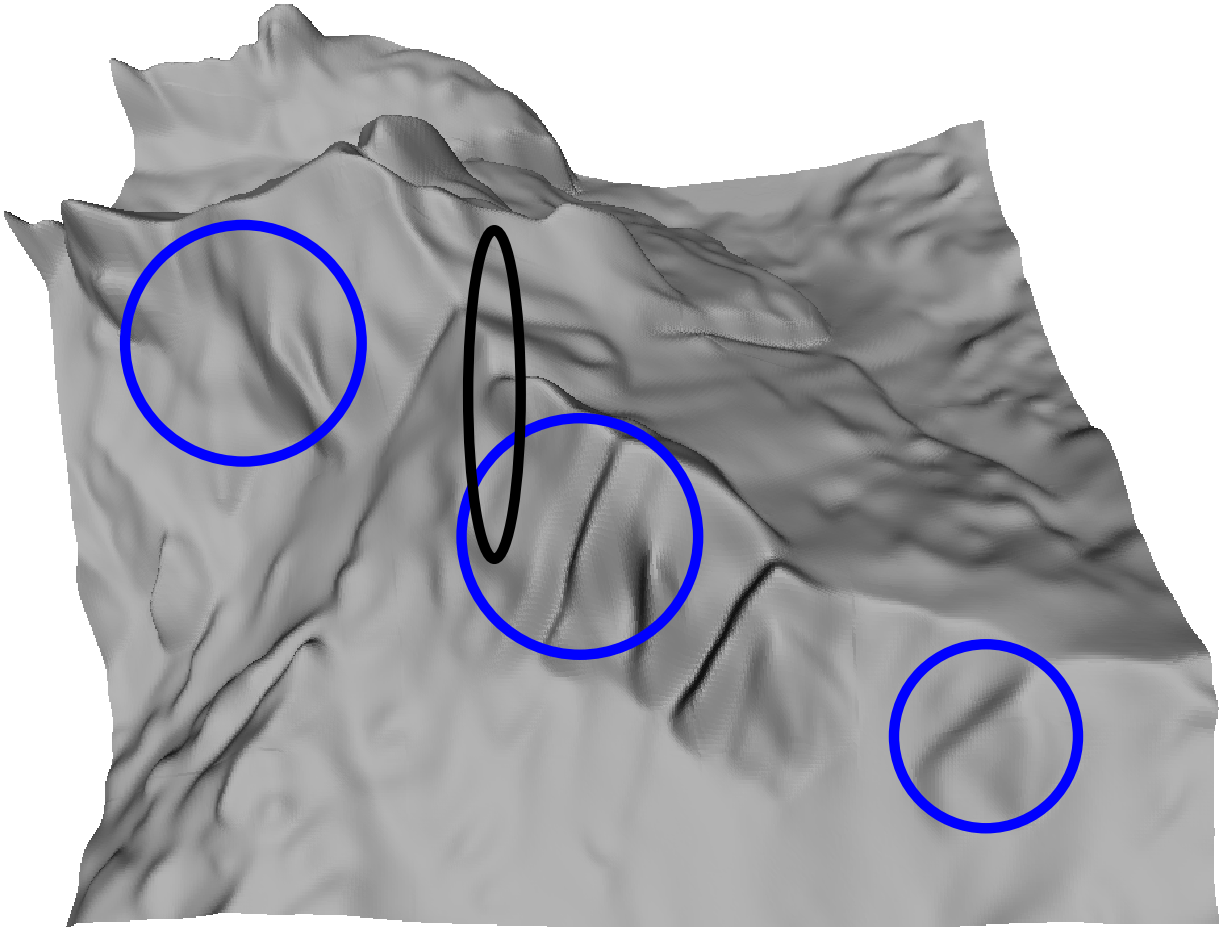,width=2.8cm}}
  \centerline{DSRFO (Ours)}\medskip
\end{minipage}
\begin{minipage}[b]{0.16\linewidth}
  \centering
\centerline{\epsfig{figure=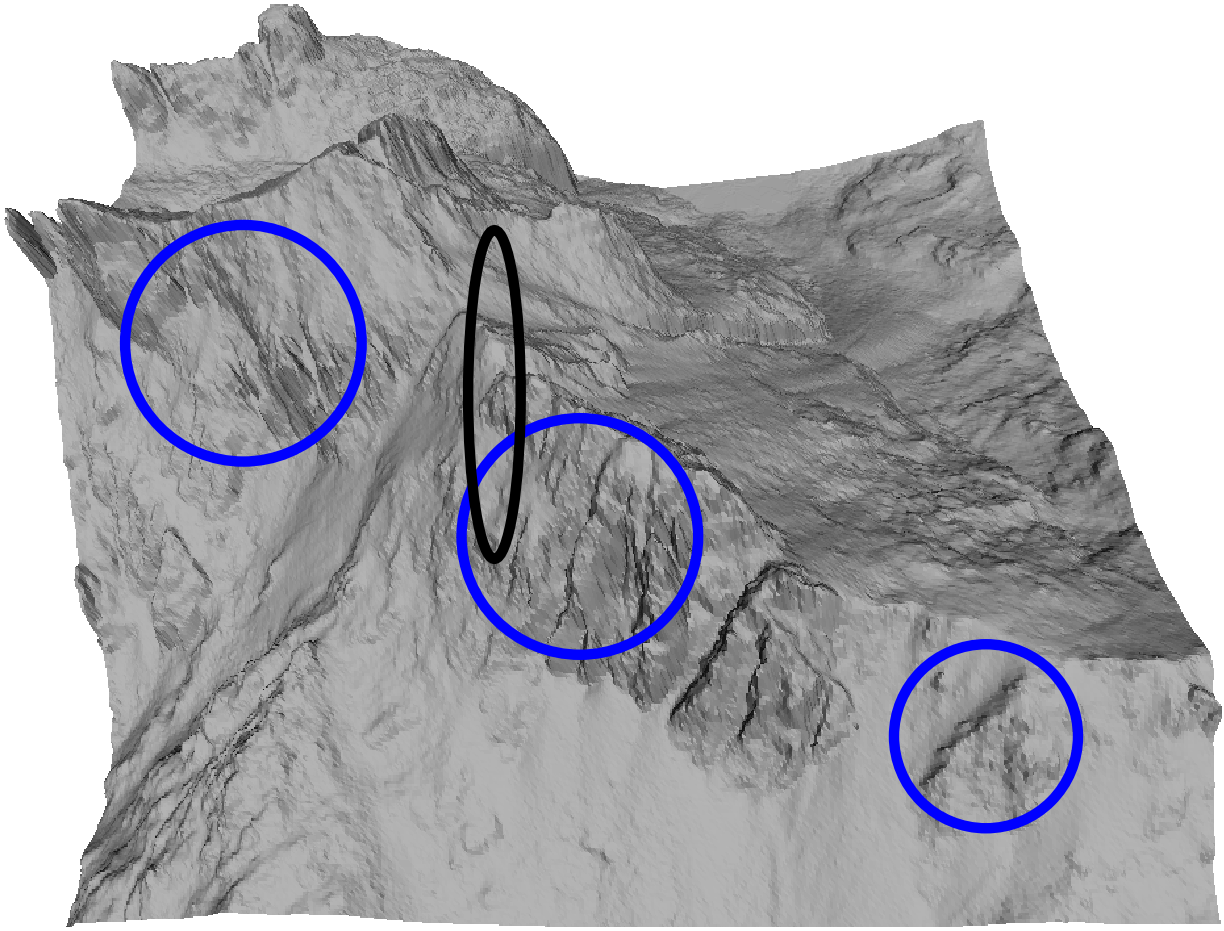,width=2.8cm}}
  \centerline{Ground Truth}\medskip
\end{minipage}
\caption{Qualitative results (best visualized in colour and zoomed-in). Blue circles show areas with interesting details. Ellipse in black shows the undersired articulation at the boundary areas for other methods but is well reconstructed in proposed DSRFO.
}
\label{fig:res}
\end{figure*}

\begin{figure}[t]
\includegraphics[width=0.4\textwidth]{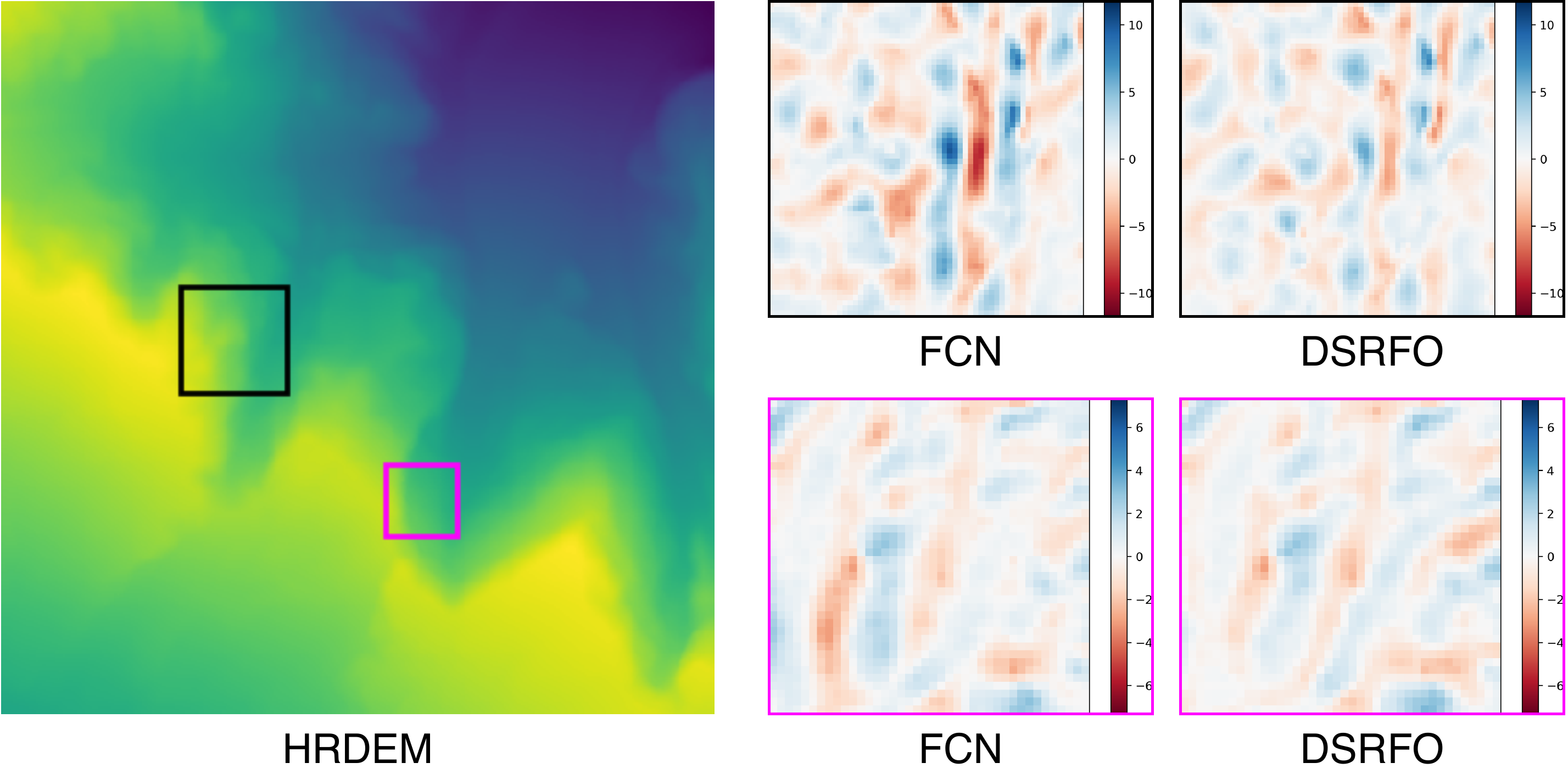}
\centering
\caption{Prediction errors displayed at patches. Red and blue intensities denote prediction below (-ve error) and above (+ve error) the actual terrain, respectively.}
\label{fig:grid}
\vspace{-0.5cm}
\end{figure}

\subsection{Comparison with state-of-art method}
We demonstrate the effectiveness of the proposed DSRFB network in comparison with the performance of  \cite{argudo2018terrain} referred as FCN and FCND (without RGB), in Table~\ref{rmse_res}. Even without using RGB channel DSRFB is able to generate high resolution DEM with nearly same accuracy. Our method performs better in regions of Dürrenstein and Monte Magro, where the aerial images respectively show them as snow-covered and with dense vegetation, indicating that method \cite{argudo2018terrain} may mislead in varying terrain landscapes. 
However, DSRFB network performs consistently and is more stable across all regions.
Further, our variant DSRFO with multiple estimations on tile boundaries performs even better and has the best performance for Dürrenstein and Monte Magro regions. Fig.~\ref{fig:grid} shows prediction errors for FCN and DSRFO on Dürrenstein region.
\section{Conclusion}
In this work, a super resolution based on feedback neural network is presented which effectively helped to enhance lower resolution terrain (LRDEM) to a higher resolution (HRDEM) without any additional input. While our method performs similar to the state of the art, the minimal input that it uses should enable better uses of DEMs. 
It can be used as a quick DEM pre-processor in terrain analysis applications. Further efforts may be needed to improve the learning from not just LRDEMs but also additional cues like in sketches or breaklines and spot heights for more enriched and feature aware terrains.


\bibliographystyle{IEEEbib}
\bibliography{refs.bib}

\end{document}